\documentclass[conference]{IEEEtran}
\usepackage{amssymb}
\usepackage{cite}
\usepackage{url}
\usepackage{amsmath}
\usepackage{xcolor}
\usepackage{csquotes}
\MakeOuterQuote{"}
\usepackage{graphicx}
\usepackage{caption}
\usepackage{subcaption}
\usepackage{times}
  
\usepackage{enumitem}
\usepackage{amsfonts}
\usepackage{xpatch}

\usepackage{scrextend}

\usepackage{algorithm}
\usepackage[noend]{algpseudocode}

\usepackage{arydshln}

\usepackage{booktabs}

\usepackage{wrapfig}

\newcounter{mybibstartvalue}
\usepackage{commath}
\usepackage{epstopdf}
\usepackage{graphicx}
\usepackage{blkarray, bigstrut} %
\usepackage{adjustbox}
\usepackage[hidelinks]{hyperref}
\usepackage[acronym]{glossaries}
\usepackage{caption}
\usepackage{amsmath, bm}
\begin{document}
	\title{Dynamic Unicast-Multicast Scheduling for Age-Optimal Information Dissemination in Vehicular Networks\vspace{-5mm}}
\author{\IEEEauthorblockN{Ahmed Al-Habob$\dagger$,  Hina Tabassum$\dagger$, Omer Waqar$\ddagger$}
	\IEEEauthorblockA{${^\dagger}$Department of Electrical Engineering and Computer Science, York University, Toronto, ON, Canada      \\
		$^\ddagger$ Department of Engineering, Thompson Rivers University, Kamloops, BC, Canada  \\
		Email: \{{alhabob, hinat}\}@yorku.ca  \{{owaqar}\}@tru.ca
}}
	
\raggedbottom		\maketitle	
	\IEEEpeerreviewmaketitle
	
	\begin{abstract}
	 This paper investigates the problem of minimizing the age-of-information (AoI) and transmit power consumption in a vehicular network, where a   roadside unit (RSU)  provides timely updates about a set of physical processes to vehicles. Each vehicle is interested in maintaining the freshness of its information status about one or more physical processes. A framework is proposed to  optimize the decisions to unicast, multicast, broadcast, or not transmit updates to vehicles as well as power allocations to minimize the  AoI and the RSU's power consumption over a time horizon. The formulated problem is a mixed-integer nonlinear programming problem (MINLP), thus a global optimal solution is difficult to achieve. In this context, we first develop an ant colony optimization (ACO) solution which provides near-optimal performance and thus serves as an efficient benchmark. Then, for real-time implementation, we develop a deep reinforcement learning (DRL) framework that captures the vehicles' demands and channel conditions in the state space and assigns processes to vehicles through dynamic unicast-multicast scheduling actions.  Complexity analysis of the proposed algorithms is presented. Simulation results depict interesting trade-offs between AoI  and power consumption as a function of the network parameters. 
	\end{abstract}
	
\section{Introduction}

With the increasing diversity of vehicular applications that require real-time information updates, such as   blind spot/lane change   and forward collision warnings, vehicular communications   over  
the upcoming 6G mobile networks become time-critical, and thus,  fresh   updates are of   high importance \cite{9779322}. While the conventional communication latency and throughput are effective metrics to evaluate the performance  of the vehicular communication networks, these  
metrics do not capture the information freshness which is critical to obtain the real-time knowledge about the location, orientation, and speed of the  vehicles. To this end, the age-of-information (AoI) is emerging as a useful metric to quantify the  freshness of the information while taking into account  the transmission latency, update generation time,  and inter-update time interval. Specifically, AoI is defined as the elapsed time between the  received information at the destination and the time when it was generated at the source \cite{6195689}. It should be noted that the inter-update time ---which is  a scheduling parameter--- is a crucial parameter in the AoI \cite{6195689}, and thus  optimizing AoI is totally different from optimizing other metrics such as the
 throughput and latency.

Moreover, the dramatic upsurge in the number of vehicles requires  
the roadside infrastructures---such as roadside
units (RSUs)---to serve more vehicles 
simultaneously and  support their time-critical update requirements. In this context, unicast and   multicast transmissions   are typically considered to transmit independent data messages  of interest for a user and    a group
of users, respectively. For instance, in \cite{8412246},   the authors considered maximizing the spectral efficiency by optimizing the downlink training and transmit power allocations. However, the unicast-multicast transmissions are predefined such that a user 
 is either  a unicast  or belongs to a group of the multicast groups.   Another predefined unicast-multicast transmission scenario was considered in \cite{8417939}, in which     each   user receives a private
  message and a common message is broadcasted to all users. The authors maximized the desired effective channel gain by designing the unicast power allocation and multicast beamformers. In  \cite{8828094}, the energy efficiency was maximized while considering a predefined  unicast-multicast scenario with simultaneous wireless information and
power transfer. 
  
None of the aforementioned research works considered \textit{optimizing a unicast, multicast, broadcast transmission scheduling}, while\textit{ minimizing AoI}. Note that the consideration of minimizing a time-dependent metric, such as  AoI necessitates a dynamic transmission scheduling  over the time horizon.
This paper develops a dynamic transmission scheduling and power allocation framework, in which at each time slot a vehicle  receives either unicast, multicast, or broadcast message from the RSU with optimized power allocations. The main contributions of this paper are summarized as follows:
\begin{itemize}
    \item We consider  minimizing both the AoI at the vehicles and the RSU's power expenditure, while optimizing the unicast-multicast scheduling decisions and their corresponding power allocations.    The two objectives are coupled in a conflicting manner, due to the transmit power allocations. Therefore, we formulate a multi-objective optimization problem for the two conflicting objectives.
    \item We develop a metaheuristic  solution based on the ant colony optimization (ACO)  to solve the  optimization problem, which provides a  near-optimal solution.  
    \item A computationally efficient solution for the real-time implementation purposes is also developed using
     deep reinforcement learning (DRL)  model, which captures the vehicles' demands and the channel conditions in the state space and assigns processes to vehicles through dynamic unicast-multicast scheduling actions.
     \item Complexity analysis of the proposed algorithms is presented. Simulation results demonstrate interesting trade-offs between AoI  and power consumption as a function of system parameters.    
     
\end{itemize}

The remainder of this paper is organized as follows.
Section~\ref{Sys} presents the  system model. The performance metrics and problem formulation are discussed in Section~\ref{Key}.      Section \ref{Sol} presents the ACO and DRL solutions. Section~\ref{SimSec}  illustrates
simulation results  and Section~\ref{Con} concludes the paper.

 \section{System Model and Assumptions}\label{Sys}
The considered system  consists of a  set $\mathcal{V} = \{v_i\}_{i=1}^{V}$ of  $ V $ vehicles supported by an   RSU  that   disseminates  timely
status updates  to  the vehicles.  The RSU is equipped with
a   uniform linear array  of $ N $ antennas. A multi-modal data dissemination scenario is considered, in which the RSU  is capable of providing timely
status updates  about a set $\mathcal{F} = \{f_l\}_{l=1}^{F}$ of $ F $ physical processes. The payload size of  an update     is   $L$ bits. Each vehicle is interested in   maintaining   
freshness of its information
status about a subset of    processes $ \mathcal{R}_i \subseteq \mathcal{F} $. To represent the information demands of the   vehicles, we define $ \bm{R}=[r_{il}]_{\footnotesize{ V\times F}} $   such that \vspace{-.01cm}
\begin{equation}
	r_{il} = \begin{cases} 1, & \mbox{if vehicle } i  ~\mbox{is interested  in process}~ l,  \\ 
		0, & \mbox{otherwise}. \end{cases}
\end{equation}
The time is divided into $ T $ time slots each of duration $ \delta $. Let $ \psi_0=\{x_0,y_0\} $  be the coordinates of the RSU and $ \psi^{(t)}_i=\{x^{(t)}_i,y^{(t)}_i\} $ be the coordinates of vehicle $ i $ at time slot $ t $. The angle of vehicle $ i $ relative to the RSU at time slot $ t $ can be expressed as 
	$\phi_i^{(t)}=\arccos \frac{x^{(t)}_i-x_0}{\ell_i^{(t)}}$, 
where $ \ell_i^{(t)} = \lVert\psi^{(t)}_i-\psi_0\rVert $ is the distance between the   vehicle $ i $ and the RSU. Let us define the process-vehicle assignment decision variable $ \bm{\eta}^{(t)}=[\eta_{il}^{(t)} ]_{\footnotesize{ V\times F}} $,  such that
\begin{equation}
	\eta_{il}^{(t)} \!\!=\!\! \begin{cases} 1, & \!\!\!\mbox{if the update of } f_l  ~\mbox{is assigned to}~ v_i ~\mbox{at time slot}~t,  \\ 
		0, & \!\!\!\mbox{otherwise}. \end{cases} \nonumber
\end{equation}    
It is worth noting that $ \sum_{i=1}^{V}\eta_{il}^{(t)}=1 $ implies that the update  of $ f_l $ is unicasted to  a single vehicle with $ \eta_{il}^{(t)}=1 $, $ \sum_{i=1}^{V}\eta_{il}^{(t)}=V'< V $ implies that the update of $ f_l $ is multicasted to   a group of vehicles with $ \eta_{il}^{(t)}=1 $.  Finally, $ \sum_{i=1}^{V}\eta_{il}^{(t)}= V $ implies that   the update message of $ f_l $ is broadcasted to  all  vehicles  and $ \sum_{i=1}^{V}\eta_{il}^{(t)}=0 $ implies that the information of process $ f_l $ is not transmitted to any vehicle at time slot $t$.  
The communication channel between the RSU and vehicle $i$ at time slot $t$ is modeled as follows:
\begin{equation}
	\mbox{\textbf{h}}_i^{(t)} = \sqrt{\frac{c_0}{4\pi f_c\ell_i^{(t)^2}}}\mbox{\textbf{a}}^H(\phi_i^{(t)}) e^{j2\pi \varrho_i^{(t)}},
\end{equation} 
where $ f_c $ is the carrier frequency,  $ c_0 $ is the speed of light, $H$ denotes the Hermitian transpose  of   $\mbox{\textbf{a}}$, and  $ \varrho_i^{(t)} $ is the
Doppler shift due to the movement of   vehicle $ i $ expressed as
$
	\varrho_i^{(t)}=\frac{c_i f_c\cos \phi_i^{(t)} }{c_0}, 
$
where $ c_i $ is the speed of vehicle $ i $ \cite{9557830}. Assuming a uniform linear antenna array at the RSU, the transmit array steering vector  $ \mbox{\textbf{a}}(\phi_i^{(t)} ) \in  \mathbb{C}^{N\times 1} $ (with  $ \phi_i^{(t)} $ as the   azimuth angle between the RSU and vehicle $i$ at time slot $t$) can be expressed as  follows:
 \begin{equation}
	\mbox{\textbf{a}}(\phi_i^{(t)})\! = \! [1,e^{j\pi \sin \phi_i^{(t)}}\!, e^{j 2\pi\sin \phi_i^{(t)} }\!,  \cdots\!, e^{j (N-1)\pi\sin \phi_i^{(t)} }],
\end{equation}
where $ j=\sqrt{-1} $ and the antenna spacing is $ \lambda/2 $ with $ \lambda $ as the carrier wavelength.

\section{Performance Metrics and Problem Statement}\label{Key}
\subsection{Decoding Error Probability}
To   guarantee the   vehicles' quality-of-service (QoS) requirements, the decoding error probability  of each message should be less than  a  tolerable decoding error. 
The decoding error probability   can be expressed as \cite{9369424} follows:
\begin{equation}\label{error}
	\varepsilon_i(\gamma_i^{(t)}) = \Phi\left(\sqrt{\frac{\delta_2 \omega}{\Gamma^{(t)}_i}}\left[\ln\left(1+\gamma_i^{(t)}\right)-\frac{L\ln 2}{\delta_2 \omega}\right]\right),
\end{equation}   
where $ \Phi\left(q\right)\triangleq\frac{1}{\sqrt{2\pi}}\int_{q}^{\infty}\exp(-\frac{u^2}{2})du $,  $ \Gamma^{(t)}_i\triangleq1-\frac{1}{(1+\gamma_i^{(t)})^2} $ is   the channel
dispersion, $ \gamma_i^{(t)} $ is the signal-to-interference plus-noise ratio (SINR) at vehicle $i$ at time slot $t$,   $ \omega $  is the bandwidth of the channel, and $ \delta_2\triangleq\delta-\delta_1 $ is the information transmission time,  with $ \delta_1 $ as the dedicated time to acquire the vehicles' angular parameters (i.e., location and speed).

\subsection{SINR Model with MRT Beamforming}
 The maximum ratio transmission (MRT)   
	beamforming scheme is considered, in which the asymptotically optimal
	beamformer vector for the vehicles that assigned the same update message  is a linear
	combination of channels of these vehicles \cite{7417526},\cite{8412246}. Consequently, for a given  $ 	\bm{\eta}^{(t)} $, the linear combination of the channel vectors of vehicles that receive a message about $ f_l $  is expressed as    $\sum_{i=1}^{V}\eta_{il}\mbox{\textbf{h}}_i^{(t)}$.  Let $ \mbox{\textbf{p}}^{(t)} =$ $[p^{(t)}_1, \cdots, p^{(t)}_F] $ be the power allocation decision with     $p^{(t)}_l$ as the allocated power to transmit   message update of $f_l$. For    a given process-vehicle assignment    $ \bm{\eta}^{(t)} $ and power allocation decision $\mbox{\textbf{p}}^{(t)}$, the MRT beamforming vector  of message $ f_l $ (the beamforming vector of the group of vehicles that receive an update about $ f_l $) is expressed as  follows:
	
	\begin{equation}
		{\mbox{\textbf{{w}}}}_l(\bm{\eta}^{(t)},\mbox{\textbf{p}}^{(t)})=	\sum_{i=1}^{V}\eta_{il}\frac{\sqrt{{p}_l^{(t)}}\mbox{\textbf{h}}_i^{(t)}}{\sqrt{N\chi^{(t)}_{i}\tilde{\xi}_{l}}},
	\end{equation}
	where $ \chi^{(t)}_{i}= {\frac{c_0}{4\pi f_c\ell_i^{(t)^2}}} e^{-j2\pi \varrho_i^{(t)}} $ is the   large-scale channel attenuation of vehicle $i$ and  $ \tilde{\xi}_{l} $ is a normalization factor \cite{8412246,8125756}. 
The SINR at vehicle $i$ can thus be expressed as follows:

	\begin{equation}\label{SNR_SDMA_20}
		\gamma_i^{(t)}(\bm{\eta}^{(t)},\mbox{\textbf{p}}^{(t)})= \frac{ |\mbox{\textbf{h}}_i^{(t)^H}\mbox{\textbf{w}}_l^{(t)}(\bm{\eta}^{(t)},\mbox{\textbf{p}}^{(t)})|^2}{\sum\limits_{\substack{m=1\\m\neq l }}^{F} |\mbox{\textbf{h}}_i^{(t)^H}\mbox{\textbf{w}}_m^{(t)}(\bm{\eta}^{(t)},\mbox{\textbf{p}}^{(t)})|^2+\sigma^2}.
	\end{equation}

\subsection{Age of Information}
The instantaneous AoI of the physical process  $f_l$ at vehicle $ i $  evolves according to 
\begin{equation}\label{Ins_AoI}
	\Delta_{il}^{(t)}\!(\bm{\eta}^{(t)}\!,\!\mbox{\textbf{p}}^{(t)}\!)\!=\!\! \begin{cases} \delta,&\!\!\!\!\!\!\!\!\!\!\!\!\!\!\!\!\!\!\!\!\!\!\!\!\!\mbox{if}~ \eta_{il}^{(t)}\!\!\!=\!1 ~\mbox{and}~ \varepsilon_i(\gamma_i^{(t)}(\bm{\eta}^{(t)},\mbox{\textbf{p}}^{(t)})\!)\!\leq\varepsilon^{\mbox{\scriptsize max}}_i\!,\\ 
		\Delta_{il}^{(t-1)}\!+\!\delta,& \mbox{otherwise}, \end{cases}
\end{equation}
where $ \varepsilon^{\mbox{\scriptsize max}}_i $ is the maximum allowed error probability at vehicle $i$. The time-average  AoI of $f_l$ at vehicle  $ i $ over $ T $ time slots is
	 $  \bar{\Delta}_{{il}}(\bm{\eta}^{(t)},\mbox{\textbf{p}}^{(t)})\triangleq\mathbb{E}_{T}\!\!\left[\Delta_{il}^{(t)}(\bm{\eta}^{(t)},\mbox{\textbf{p}}^{(t)})\right]=$ $ \frac{1}{T}\sum_{t=1}^{T}	\Delta_{il}^{(t)}(\bm{\eta}^{(t)},\mbox{\textbf{p}}^{(t)}) $. Consequently, the total time-average  AoI    can be expressed as follows:
	 \begin{equation}\label{Avv_AoI}
	 	\begin{split}
	 	\bar{\Delta}(\bm{\eta}^{(t)},\mbox{\textbf{p}}^{(t)})&=\sum_{i=1}^{V}\sum_{l=1}^{F} r_{il}\bar{\Delta}_{{il}}(\bm{\eta}^{(t)},\mbox{\textbf{p}}^{(t)})\\
	 	&=\frac{1}{T}\sum_{i=1}^{V}\sum_{t=1}^{T}\sum_{l=1}^{F} r_{il}	\Delta_{il}^{(t)}(\bm{\eta}^{(t)},\mbox{\textbf{p}}^{(t)}). 
	 	\end{split}
	 \end{equation}
Note that the maximum value of $ \bar{\Delta}_{{il}} $ is $ \delta(T+1)/2 $, which corresponds the case of no update about $f_l$ is received at vehicle $ i $ over the $ T $ time slots. Thus, the maximum value (upper bound) of the total time-average  AoI $ \bar{\Delta}^{\mbox{\scriptsize max}} $ can be expressed as:
\begin{equation}\label{max_AoI}
	\bar{\Delta}^{\mbox{\scriptsize max}} = \frac{\delta (T+1)}{2}\sum_{i=1}^{V}\sum_{l=1}^{F}r_{il},
\end{equation}   
which corresponds the case of no update is received by any vehicle during the $ T $ time slots. The minimum value (lower bound) of the total time-average  AoI $ \bar{\Delta}^{\mbox{\scriptsize min}} $ corresponds the case of updating each vehicle  in each time slot. Keeping in mind that a vehicle $ i $ is interested in $ |\mathcal{R}_i|=\sum_{l=1}^{F}r_{il} $ processes and the best option is to alternate the updates between the  $ |\mathcal{R}_i| $ processes, $ \bar{\Delta}^{\mbox{\scriptsize min}} $
  can be expressed as\footnote{The expression in \eqref{min_AoI}  is valid for $ T > |\mathcal{R}_i|+1  $; for the special case $ T < |\mathcal{R}_i| -1 $, we have $ \bar{\Delta}^{\mbox{\scriptsize min}} =\sum_{i=1}^{V} \frac{\delta}{T} \left[|\mathcal{R}_i| \frac{T(T+1)}{2} -\sum_{r=1}^{T}\frac{r(r+1)}{2}\right] $; both expressions are equivalent for $|\mathcal{R}_i|-1\leq T\leq|\mathcal{R}_i|+1$.}:   
\begin{equation}\label{min_AoI}
	\bar{\Delta}^{\mbox{\scriptsize min}} =\sum_{i=1}^{V} \frac{\delta}{T}\left[ T\frac{|\mathcal{R}_i|(|\mathcal{R}_i|+1)}{2}- \sum_{r=1}^{|\mathcal{R}_i|-1}\frac{r(r+1)}{2}\right].
\end{equation}   
  
  \begin{figure}[t]
  	\begin{center}
  		\includegraphics[width=.99\linewidth]{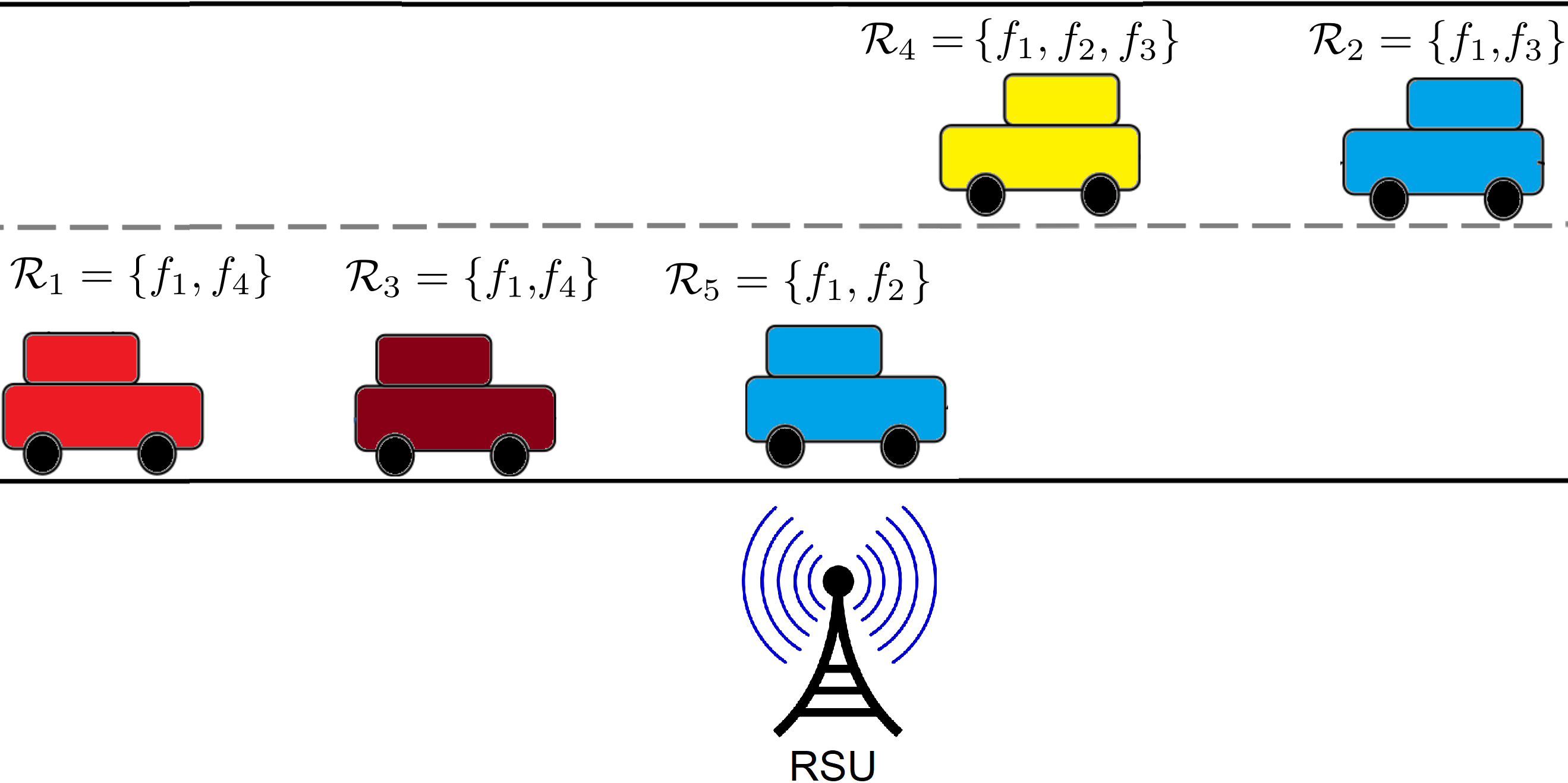}
  		\caption{Illustrative example with $V=5$ vehicles and  $F=4$   processes.}\label{Mot}
  	\end{center}
  \end{figure}
  
\subsection{Significance of Process-Vehicle Assignment and Power Allocation - A Toy Example}\label{Mot_example}
Let us consider $V=5$ vehicles, where each vehicle is interested in a subset of  $\mathcal{F} =\{f_1, f_2, f_3, f_4\}$ processes with demands as illustrated in Fig. \ref{Mot}. The associated information demand matrix $\mathbf{R}$   can be represented as follows:   
\begin{equation}
	\mathbf{R}=[r_{il}]_{\footnotesize{ 5\times 4}}=\begin{blockarray}{ccccc}
		f_1 & f_2 & f_3 & f_4  \\
		\begin{block}{[cccc]c}
			1 & 0 & 0 & 1 & v_1\\
			1 & 0 & 1 & 0 &  v_2 \\
			1 & 0 & 0 & 1 &  v_3 \\
			1 & 1 & 1 & 0 & v_4 \\
			1 & 1 & 0 & 0 & v_5 \\
		\end{block}
	\end{blockarray}.
\end{equation}
Let us assume that the observation interval is $T=7$ time slots and  other parameters are as listed in Table \ref{TableResults}. According to \eqref{max_AoI} and \eqref{min_AoI}, the maximum and minimum time-average AoI at the vehicles can be computed  $\bar{\Delta}^{\mbox{\scriptsize max}} = 308/7= 44$s  and $ \bar{\Delta}^{\mbox{\scriptsize min}} = 108/7 \approx 15.4$s, respectively. 
Let us assume that the RSU assigns the processes to vehicles and allocates the power at random, the time-average   AoI will  vary around $ 210/7 \approx 30$s and the power consumption will be around $0.5$W. By   examining    all   possible  process-vehicle assignment  and power allocation decisions  with objective of minimizing both AoI and power consumption,   the time-average   AoI can be reduced to  $ 124/7 \approx 17.7$s while the time-average consumed power   is $0.18$W if the following decisions are performed. 
At $ t=1 $ no message is transmitted. At $t=2$, an update about $ f_1 $ is broadcasted to all vehicles. At $t=3$, an update about $ f_3 $ is unicasted to $v_2$ and updates about $ f_4 $ and $ f_2 $  are multicasted to $\{v_1, v_3\} $ and $\{v_4, v_5\} $, respectively,   and so on. 
The question is how to select optimal  scheduling and power allocation decisions at each time slot for arbitrary number of physical processes and  vehicles with different demands. This motivates the problem formulation in the following.

\subsection{Problem Formulation}\label{Prob}
We consider minimizing the time-average AoI of each process at the vehicles   as well as the time-average   power consumption at RSU.  Keeping in mind   the trade-off between   these two objectives and the fact that   they  have different units, ranges, and orders of magnitude, they
should be normalized such that they have similar ranges \cite{marler2004survey}. Consequently,  we define    a    multi-objective weighted sum  utility function    as:
\begin{equation}\label{Obj}
	\begin{split}
		{O}(\bm{\eta}^{(t)}\!,\!\mbox{\textbf{p}}^{(t)})\!=& \zeta\frac{\bar{\Delta}(\bm{\eta}^{(t)}\!,\mbox{\textbf{p}}^{(t)}\!) \!-\!\bar{\Delta}^{\mbox{\scriptsize min}}}{\bar{\Delta}^{\mbox{\scriptsize max}}-\bar{\Delta}^{\mbox{\scriptsize min}}}\!+\!(1\!-\!\zeta)\frac{\bar{{p}}^{(t)}\!-\!P^{\mbox{\scriptsize min}}}{P^{\mbox{\scriptsize max}}\!-\!P^{\mbox{\scriptsize min}}},
	\end{split}
\end{equation}
 where  the time-average power consumption can be given as $ \bar{{p}}^{(t)} = \frac{1}{T}\sum_{t=1}^{T}\sum_{l=1}^{F}p^{(t)}_l $,  $ 0 < \zeta \leq 1$ is a relative weight to give preference to minimize the AoI or the power, $ \bar{\Delta}^{\mbox{\scriptsize max}} $ and $ \bar{\Delta}^{\mbox{\scriptsize min}} $ are the maximum and minimum total time-average AoI as expressed in \eqref{max_AoI} and \eqref{min_AoI}, respectively,   $ P^{\mbox{\scriptsize max}} $ is the maximum transmission power of the RSU, and $ P^{\mbox{\scriptsize min}}=0 $.   The   optimization problem is thus formulated  as follows: 
 
\begin{subequations}\label{P1}
	\begin{alignat}{2}
		\mbox{\textbf{P1}}~	&\!\min_{\bm{\eta}^{(t)},\mbox{\textbf{p}}^{(t)}}&& {O}(\bm{\eta}^{(t)},\mbox{\textbf{p}}^{(t)}) \label{OPP1}\\
		&\mbox{s.t.}&& {\sum_{l=1}^{F}p_l^{(t)} \leq P^{\mbox{\scriptsize max}}},\label{con1}\\
		&&~~   & \sum_{l=1}^{F} \eta_{il}^{(t)} \leq 1, ~ \forall~ v_i \in \mathcal{V},\label{con3}\\
				&&~~   &   \eta_{il}^{(t)} r_{il} = \eta_{il}^{(t)}, ~ \forall~ 1\leq i\leq V, 1\leq l\leq F, \label{con4}\\
					&&~~   & {p_l^{(t)}\geq 0, ~ \forall~ f_l \in \mathcal{F}},\label{con2}\\
		&                  & ~~     & \eta_{il}^{(t)}\in \{0, 1\},  ~ \forall v_i  \in \mathcal{V}, f_l  \in \mathcal{F}. \label{con5}
	\end{alignat}
\end{subequations}
Constraint \eqref{con1}  guarantees that the allocated power is less than the maximum transmission power of the RSU. Constraint \eqref{con3}  guarantees that at most one process is assigned to each vehicle at each time slot. 
Keeping in mind that $ \bm{R} $ and $  \bm{\eta}^{(t)} $ are binary variables, \eqref{con4} guarantees that if a vehicle $ i $ is not interested in process $l$ (i.e., $ r_{il}=0 $) then process $l$ will not be assigned to vehicle $i$ (i.e., $ \eta_{il}^{(t)}  $ should be $0$). 

The optimization problem in \eqref{P1} is a mixed-integer non-linear programming (MINLP) problem, where the discontinuity in the objective function comes from (8), the non-linearity comes from (6),   and the integer constraint arises from (14f).  

	\section{Proposed Age-Optimal Solutions}\label{Sol}
First, it is important to note that from \eqref{Ins_AoI},    the AoI of process $f_l$ at vehicle $i$ can be minimized if both $\eta_{il}^{(t)}=1$ and  $\varepsilon_i(\gamma_i^{(t)}(\bm{\eta}^{(t)},\mbox{\textbf{p}}^{(t)})\!)\leq\varepsilon^{\mbox{\scriptsize max}}_i$. To ensure that $\varepsilon_i(\gamma_i^{(t)}(\bm{\eta}^{(t)},\mbox{\textbf{p}}^{(t)})\!)\leq\varepsilon^{\mbox{\scriptsize max}}_i$, we denote
 $ \hat\gamma_i $ be the SINR at vehicle $v_i$ that ensures    $\varepsilon_i(\hat \gamma_i)=\varepsilon^{\mbox{\scriptsize max}}_i$ which can be found by solving \eqref{error} numerically. Then, from \eqref{error} and \eqref{SNR_SDMA_20}, the AoI of the process $f_l$ at vehicle $v_i$ can be minimized if both $\eta_{il}^{(t)}=1$ and the following inequality are satisfied, i.e., \vspace{-.31cm}
 \begin{equation}\label{Const}
 \frac{ {p}_l^{(t)}|\mbox{\textbf{h}}_i^{(t)^H}\mbox{\textbf{w}}_l^{(t)}(\bm{\eta}^{(t)})|^2}{\sum\limits_{\substack{m=1\\m\neq l }}^{F}{p}_m^{(t)} |\mbox{\textbf{h}}_i^{(t)^H}\mbox{\textbf{w}}_m^{(t)}(\bm{\eta}^{(t)})|^2 +\sigma^2}\geq \hat \gamma_i.    
 \end{equation}
 \vspace{-.31cm}

 Consequently, a solution for   
 the optimization problem in \eqref{P1} can be obtained by solving the following  optimization problem:
 \begin{subequations}\label{P3}
 	\begin{alignat}{2}
 		& \textbf{P2} \min_{\bm{\eta}^{(t)}{, \mbox{\textbf{p}}^{(t)}}}&&~ {O}(\bm{\eta}^{(t)},\mbox{\textbf{p}}^{(t)}_{\bm{\eta}})  \nonumber\\
 		&\mbox{s.t.}&& \eqref{con3}, \eqref{con4}, ~\mbox{and}~\eqref{con5}, \nonumber\
 	\end{alignat}
 \end{subequations}
where $ \mbox{\textbf{p}}^{(t)}_{\bm{\eta}}\!=[{p}^{(t)}_{\bm{\eta}_1}, \cdots, {p}^{(t)}_{\bm{\eta}_F}] $ is the solution of the following linear programming problem: 
\vspace{-.21cm}
 \begin{subequations}\label{PowerP1}
	\begin{alignat}{2}
	&\mbox{\textbf{p}}^{(t)}_{\bm{\eta}}\!=	\!\min_{\mbox{\textbf{p}}^{(t)}} \sum_{l=1}^{F}p_l^{(t)}  \nonumber\\
		&\mbox{s.t.} 
		\frac{ \eta_{il}^{(t)}{p}_l^{(t)}|\mbox{\textbf{h}}_i^{(t)^H}\!\!\mbox{\textbf{w}}_l^{(t)}(\bm{\eta}^{(t)})|^2}{\sum\limits_{\substack{m=1\\m\neq l }}^{F}\!{p}_m^{(t)} |\mbox{\textbf{h}}_i^{(t)^H}\!\!\mbox{\textbf{w}}_m^{(t)}(\bm{\eta}^{(t)})|^2 \!+\!\sigma^2\!}\!\!\geq\! \eta_{il}^{(t)} \hat\gamma_i, 
	                  ~~ \forall  f_l ,   v_i,\label{conPower1}
 \\
		& ~~ \eqref{con1} ~\mbox{and}~ \eqref{con2}	\nonumber	\end{alignat}
\end{subequations}
It is worth noting that for a given $  \bm{\eta}^{(t)} $,  constraint \eqref{conPower1} guarantees  that the allocated power minimizes the AoI. If \eqref{PowerP1} is infeasible for a given $  \bm{\eta}^{(t)} $, then  $  \bm{\eta}^{(t)} $ does not minimize the objective function and will be discarded.  The 
 strategy of the following ACO and DRL solution approaches can be summarized as follows. Search for an optimized $  \bm{\eta}^{(t)} $ by exploring its search space and for each candidate  $  \bm{\eta}^{(t)} $ find the corresponding best power allocation by solving \eqref{PowerP1}. The fitness of a candidate  $  \bm{\eta}^{(t)} $ is reflected by its ability to minimize the objective function with a feasible power allocation.


%
%

\subsection{Ant Colony Optimization (ACO)
	 }
 The optimization problem in \eqref{P3}   can be solved 
	 by enumerating all the feasible decisions. Such exhaustive search approach is computationally inefficient, which motivates  designing a metaheuristic solution based on the ACO        for rapid
	 discovery of good solutions and guarantee convergence \cite[Ch.4.3]{ACObook}. 
	  In the proposed ACO algorithm, a colony of  $A$ ants  collaborate to solve $\textbf{ {P1}}$. Each ant   $ a \in A $   travels  a   tour of $T$ steps. In each step, it assigns a process for each vehicle.
The probability that     ant $ a $ assigns process $ f_l $ to vehicle $ v_i $     and the probability it does not  assign a process  to vehicle $ v_i $  at time  $t$  are    obtained as:  
\begin{equation}\label{bid}
	\begin{split}
		\bar{\pi}_{il}^{\left(t\right)}=\frac{(\bar{\tau}^{(t)}_{il})^{\iota_1}(\bar{\varrho}^{(t)}_{il})^{\iota_2}}{1+\sum_{l=1}^{F}(\bar{\tau}^{(t)}_{il})^{\iota_1}(\bar{\varrho}^{(t)}_{il})^{\iota_2}},~
		\bar{\pi}_{i0}^{\left(t\right)}= 1-\sum_{l=1}^{F}\bar{\pi}_{il}^{\left(t\right)},
	\end{split}
\end{equation}
respectively, where  $ \bar{\tau}^{(t)}_{ir} $   is the trail pheromone and $ \bar{\varrho}^{(t)}_{il}$ is the   attractiveness   of assigning process $f_l$ to vehicle $v_i$. The latter is  set to be
\begin{equation}
    \bar{\varrho}^{(t)}_{il} = 
		\frac{r_{il}\Delta_{il}^{(t)}}{t},
\end{equation}
to   give higher attractiveness to assigning the process with high AoI and of interest to  $v_i$.  { The parameters $\iota_1 $ and $\iota_2$   control the influence of the pheromone and attractiveness, respectively.}  At each step, ant $a$ obtains $  \bm{\eta}^{(t,a)} $ based on \eqref{bid} and obtain the power allocation $ \mbox{\textbf{p}}^{(t,a)} $  by  solving \eqref{PowerP1}. Only ants  that  obtain the highest and second-highest $O(\bm{\eta}^{(t,a)},\mbox{\textbf{p}}^{(t,a)}) $ $ \forall a \in A $   deposit their pheromone \cite{doerner2004pareto}.
The 
pheromone is   updated as follows:
\begin{equation} \label{deposit}
		\bar{\tau}_{il}^{(t)}\leftarrow\left(1-\kappa\right)\bar{\tau}_{il}^{(t)}+ {\eta}_{il}^{(t,a)}\nabla{\tau}^{(a)}, 
\end{equation}
where $ \kappa $  is the pheromone evaporation coefficient and $ \Delta{\tau}^{(a)} $ is the deposit pheromone, which is expressed as 
\begin{equation}
	 \nabla{\tau}^{(a)}\!\!=\!\begin{cases} \exp[-O(\bm{\eta}^{(t,a)},\mbox{\textbf{p}}^{(t,a)})], &\!\!\!\!\mbox{if}~  \sum\limits_{l=1}^{F}p_l^{(t,a)} \!\leq\! P^{\mbox{\scriptsize max}}\!,\\ 
		0, & \mbox{otherwise}. \end{cases}
\end{equation}
The ACO algorithm is illustrated in \textbf{Algorithm \ref{euclid1}} which  iterates until the improvement in the  best solution of the whole colony is less than a threshold $\epsilon_0$ or a maximum number of colonies $ I $  has been generated.

\begin{algorithm}
	\small 
	\caption{\small ACO algorithm for age-optimum dynamic transmission.}\label{euclid1}
	\begin{algorithmic}[1]
 \small
 	\State  {\textbf{Input} $\bm{R}$,  $\mbox{\textbf{h}}_i^{(t)}$, $\kappa$, $I$, $A$, $\epsilon_0$;}
	\State  $ O^* \leftarrow  \infty $; {$ O_p \leftarrow  0 $;}
		\State    {\textbf{while} $ I \geq 1 $ and $ |O^* -O_p|\geq \epsilon_0 $  \textbf{do}} 
		\State $ ~~ $ $O_1 \leftarrow  \infty $; $O_2 \leftarrow  \infty $; {$I=I-1$; $O_p=O^*$;}
		\State $ ~~ $ \textbf{for} $ a =1 $ to $ A $  \textbf{do}
		\State $ ~~~~~ $ \textbf{for} $ t =1 $ to $ T $  \textbf{do}
		\State $ ~~~~~~ $ Obtain  $\bm{\eta}^{(t,a)}$ using \eqref{bid}; Obtain  $\mbox{\textbf{p}}^{(t,a)}$ by solving \eqref{PowerP1};
		\State  $ ~~~~~ $  \textbf{end for} 
		\State $~~$ Evaluate $O(\bm{\eta}^{(t,a)},\mbox{\textbf{p}}^{(t,a)})$ using \eqref{Obj};
		\State  $ ~~ $  \textbf{if} $ O^* >  O(\bm{\eta}^{(t,a)},\mbox{\textbf{p}}^{(t,a)})$, 
	 $ O^* =  O(\bm{\eta}^{(t,a)},\mbox{\textbf{p}}^{(t,a)})$;
	     \textbf{end if};		
		\State  $ ~~ $  \textbf{if} $ O_1 >  O(\bm{\eta}^{(t,a)},\mbox{\textbf{p}}^{(t,a)})$,  $ a_1=a$; 
		\State  $ ~~ $  \textbf{else}  \textbf{if} $ O_2 > O(\bm{\eta}^{(t,a)},\mbox{\textbf{p}}^{(t,a)})$,  $ a_2 = a$; 
	 \textbf{end if}
		\State  $ ~ $  \textbf{end for} 
		\State  $ ~~~ $ Deposit pheromone of $ a_1 $ and $ a_2 $ using \eqref{deposit};		
		\State   \textbf{end while} 
		\State  \textbf{Return} $O^*$.
		\normalsize
	\end{algorithmic}
\end{algorithm}

{\subsection{DRL-Based Solution Approach}
In this section, we aim to design  a real-time solution approach for the optimization problem in \eqref{P1}.   In this context, we
develop a DRL model   that involves the definition of the environment state space
$ \bm{s}^{t}\in \bm{\mathcal{S}} $, the action space $ \bm{\mu}^{(t)}\in \bm{\mathcal{A}} $, and the    immediate reward function $ \rho^{(t)} $.  

\begin{itemize}
	\item The state of the environment at time slot $ t $  is given by  
	\begin{equation}
		\bm{s}^{t}\!=\!\!\left\{ \{\chi^{(t)}_i\},\{ r_{i1}\Delta_{i1}^{(t)},r_{i2}\Delta_{i2}^{(t)}, \cdots\!, r_{iF}\Delta_{iF}^{(t)}\}\right\}_{i=1}^V\!,
	\end{equation}
	which captures the   channel state and the AoI of the processes of interest for the $ V $ vehicles.
	\item The  action at time slot $ t $ is defined as $ \bm{\mu}^{(t)}=[\mu^{(t)}_i]_{\footnotesize{ V\times 1}} $, which is a  vector of integers   $ \mu^{(t)}_i \in \{0 \cup \mathcal{R}_i\} $, such that
	\begin{equation}
		\mu^{(t)}_i = \begin{cases} l,&  \mbox{implies }  f_l  ~\mbox{is assigned  to}~ v_i,  \\ 
			0, &  \mbox{implies}~v_i~\mbox{is not updated at}~t. \end{cases}
	\end{equation}   
 It is worth noticing  that $\bm{\mu}^{(t)}$ is an equivalent representation to $\bm{\eta}^{(t)}$ with a dimension  of $V\times 1$ instead of $V\times F$. 
	\item The immediate
	reward at time slot $ t $ is expressed as
\begin{equation}
		\rho^{(t)}\!\!=\!\zeta\frac{\bar{\Delta}_t(\bm{\eta}^{(t)},\mbox{\textbf{p}}^{(t)}) \!-\!\bar{\Delta}_t^{\mbox{\scriptsize min}}}{\bar{\Delta}_t^{\mbox{\scriptsize max}}-\bar{\Delta}_t^{\mbox{\scriptsize min}}}\!+\!(1\!-\zeta)\frac{\hat{\mbox{\textbf{p}}}^{(t)}-P^{\mbox{\scriptsize min}}}{P^{\mbox{\scriptsize max}}-P^{\mbox{\scriptsize min}}},
\end{equation}
where $ \hat{\mbox{\textbf{p}}}^{(t)}=\frac{1}{t}\sum_{t'=1}^{t} {\mbox{\textbf{p}}}^{(t')} $, $\bar{\Delta}_t(\bm{\eta}^{(t)},\mbox{\textbf{p}}^{(t)})$,  $ \bar{\Delta}_t^{\mbox{\scriptsize max}} $,  and $\bar{\Delta}_t^{\mbox{\scriptsize min}}$ are obtained by replacing $T$ by $t$ in \eqref{Avv_AoI}, \eqref{max_AoI}, and \eqref{min_AoI}, respectively. {The reward is set to $ -\log(\rho^{(t)}+\nu) $, where $ \nu $ is a very small number that introduced to avoid infinite  reward.}	
\end{itemize}
The DRL training algorithm is illustrated in \textbf{Algorithm~2}.  It is worth mentioning that the trained DRL agent estimates an action $\bm{\mu}^{(t)}$ for a given state, the corresponding $\bm{\eta}^{(t)}$    is utilized to find the power allocation in \eqref{PowerP1}   and evaluate the objective function in \eqref{OPP1}.

\begin{algorithm}[ht]   
	\caption{\small DRL training algorithm   for age-optimum dynamic transmission.}\label{DeepLearning}
	\begin{algorithmic}[1]\small
		\State  Initialize the  weights of the deep   network $\theta $ and the replay buffer  \textbf{B};\label{l2}
		\State  \textbf{For $ episode=1 $ to} Max no. of episodes \textbf{do}\label{l3} 
		\State  Initialize the
		environment and receive the initial state $ \bm{s}^{(1)} $; \label{l4}
		\State $~$\textbf{Repeat:} 
		\State $ ~~$With probability $ \epsilon $, select a random action $\bm{\mu}^{(t)} \in \mathcal{A}$; Otherwise, select $ \bm{\mu}^{(t)}=\arg\max\limits_{\forall \bm{\mu}^{(t)}\in \bm{\mathcal{A}}}Q(\bm{s}^{(t)},\bm{\mu}^{(t)}\mid\bm{\theta}) $;
		\State $ ~~$ Obtain $\bm{\eta}^{(t)}$; Obtain $ \mbox{\textbf{p}}^{(t)^*}  $  by solving \eqref{PowerP1}; 
		\State $ ~~$  Observe the reward  $ \rho^{(t)} $ and    next state $\bm{s}^{({t+1})}  $;  
		\State $ ~~$	Store the transition $ \{\bm{s}^{({t})},\bm{\mu}^{(t)}, \rho^{(t)},\bm{s}^{({t+1})}\}$ in \textbf{B};
	  $ t=t+1 $;
		\State	$~$\textbf{Until} terminal state $t=T$.
		\State $ ~~~$ 
		Sample a random mini-batch of $ M $ transitions from \textbf{B};		
		\State $ ~~~$ For each transitions  in $ M $ obtain $y_m$ such that  \begin{equation*}
			y_m\!=\!\! \begin{cases}\rho^{(m)},&\!\!\!\!\!\!\!\!\!\!\!\!\!\!\!\!\!\!\!\!\!\!\!\!\!\!\!\!\!\!\!\!\mbox{if }\bm{s}^{(m+1)}  ~\mbox{is a terminal state},  \\ 
				\rho^{(m)}\!+\!\gamma\max\limits_{\forall \bm{\mu}^{(m)}\in \mathcal{A}}Q(\bm{s}^{{t}},\bm{\mu}^{(m)}\mid\bm{\theta}),&\!\!\!\!\mbox{otherwise}. \end{cases}
		\end{equation*}		
		\State $ ~~~$  Update the weight of the DNN network by minimizing the loss $ L(\bm{\theta})=\frac{1}{M}\sum_{m=1}^{M}(y_{m}-Q(\bm{s}^{(m)},\bm{\mu}^{(m)}\mid\bm{\theta}))^2 $;
		\State \textbf{End for}     
		\normalsize
	\end{algorithmic}
\end{algorithm}
}

\subsection{Complexity Analysis}
Keeping in mind that a process-vehicle assignment decision $  \bm{\eta}^{(t)} $ can be represented by a vector $ \bm{\mu}^{(t)}=[\mu_{i}^{(t)} ]_{\footnotesize{ 1\times V}} $, the 
\begin{table}[ht]
	\caption{Simulation Parameters.}
	\begin{center}
		\begin{adjustbox}{width=.48\textwidth}			\begin{tabular}{|c|c||c|c||c|c|}
				\hline				\textbf{Parameter}&\textbf{Value} & \textbf{Parameter}&\textbf{Value}   & \textbf{Parameter}&\textbf{Value}    \\ \hline
				$ P^{\mbox{{\footnotesize max}}} $ 	&   $ 0 $ dB  &  $ c_i $ 	& $\sim U(10, 15)  $ m/s  \cite{9557830}  &   $ \omega $	&   $ 10 $ MHz \cite{9557830} \\ \hline    
				$ f_c $ 	&   $ 3 $ GHz  \cite{9557830}&  $ N $ 	&   $ 64 $  \cite{9557830} &  $ L $ 	&   $128 $ byte  \\ \hline   
				$  \varepsilon^{\mbox{{\footnotesize max}}} $ 	&   $ 10^-6 $ &  $ c_0 $ 	&   $ 2.99\times 10^8$ m/s & $ \sigma^2 $ 	&   $ 0.1  $   \\ \hline   					   
				$ I $	&   $ 400 $ iteration &  $ A $ 	&   $ 100$ individuals  &   $ \delta $ 	&  $ 1   $ s  \\  \hline   
				$ \kappa $	&   $ 0.1 $ &  $ \iota_1/\iota_2 $ 	&   $ 1$  &   $ {\epsilon_0 }$ 	&  {$0.01 $}    \\  \hline   
				$ \textbf{B} $	&   $ 10^4 $ &   mini-batch 	&   $ 64$  &   learning rate	&  $ 0.001 $    \\  \hline                 
			\end{tabular}
		\end{adjustbox}
	\end{center}
	\label{TableResults}
\end{table}  
 search space of \eqref{P3} over the $T$ time slots is $ \mathcal{O}((F+1)^{V}T) $. The  worst-case computational complexity of evaluating the objective function in \eqref{Obj} is  $ \mathcal{O}(VFT^2) $ operations and solving \eqref{PowerP1}  requires   $ \mathcal{O}((V+F)^{2.5}) $ operations. Consequently, the computational complexity of the exhaustive search is  $ \mathcal{O}(V^{3.5}T^3F^{V}) $. An ant agent performs $ \mathcal{O}(V F) $ operations to 
  find a process-vehicle assignment decision,   $ \mathcal{O}((V+F)^{2.5}) $ operations to allocate the power, and  $ \mathcal{O}(VFT^2) $ operations to evaluate  \eqref{Obj}. Consequently, the  worst-case computational complexity of the ACO is  $ \mathcal{O}((VF)^{4.5}T^3AI) $, where $A$ is the number of ants in the colony and $I$ is the  maximum number of colonies. Considering a scenario of $V=5$ vehicles and $F=4$ processes, the   average execution time of the exhaustive search, ACO, and DRL  solution approaches is $ 1.8 $ s, $ 0.2 $ s, and $0.01$ s, respectively. \ {Average execution time using MATLAB on an Intel(R) Core(TM) i7-3770 CPU machine      working at a clock frequency of 3.4	GHz  and 16 GB of RAM.}

\section{Simulation Results and Discussions}\label{SimSec}
This section introduces simulation results to    evaluate  the proposed framework and compare its       
performance with the random solution approach, in which the  process-vehicle assignment and power  allocations decisions  are selected at random.   
   Unless
otherwise stated, the considered     numerical values of the system  parameters and solution approaches  are listed in Table
\ref{TableResults}.

\begin{figure*}
\vspace{-3mm}
\centering
\begin{minipage}{0.301\linewidth}
\includegraphics[width=1.\linewidth]{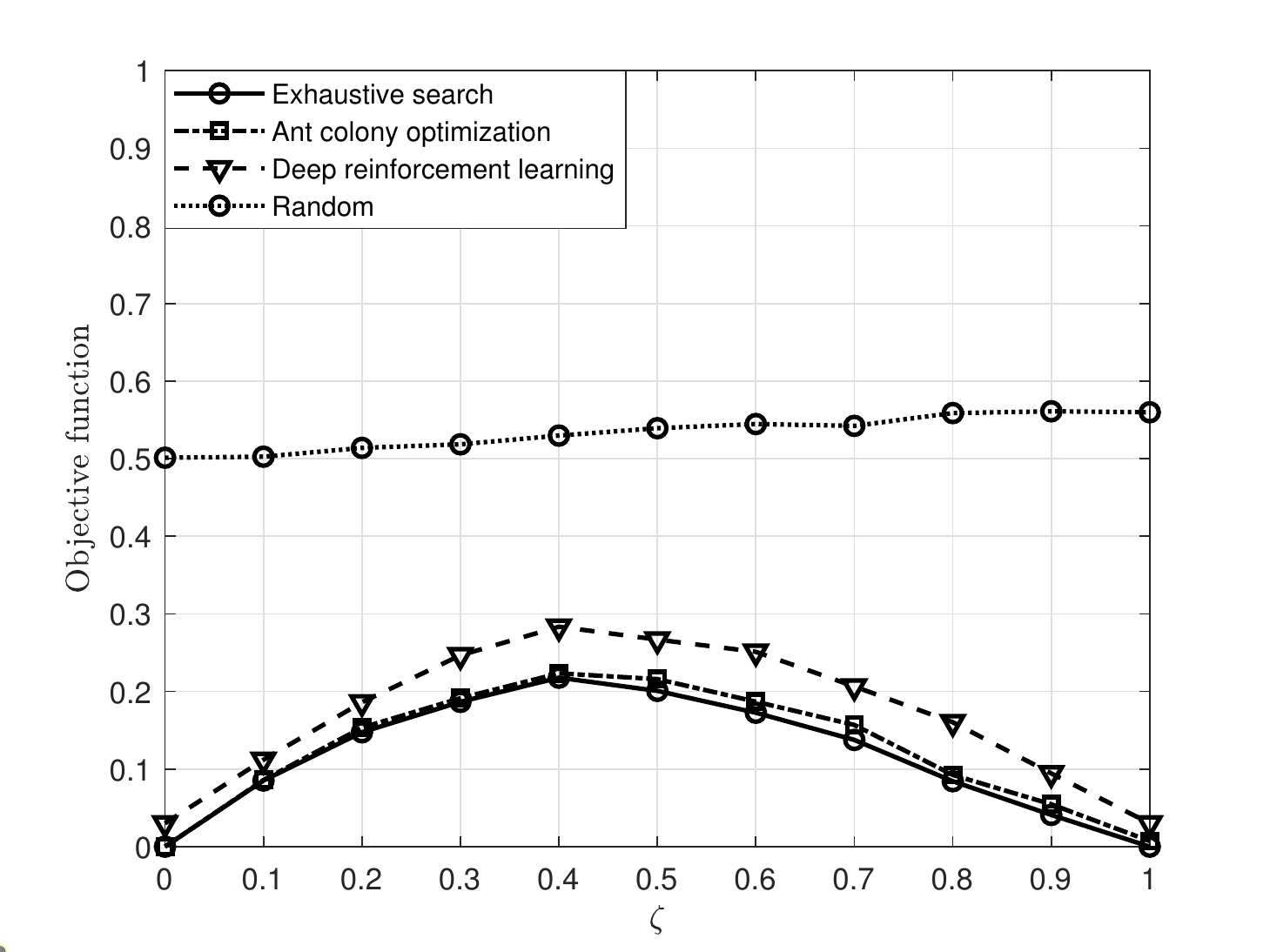}
		\caption{Objective function in \eqref{Obj}    versus  the relative weight $\zeta$ with total number of   $F=4$   process and $\abs{\mathcal{R}_i}=2$.}\label{figRes1}
\end{minipage}\hfill
\begin{minipage}{0.301\linewidth}
\includegraphics[width=1.\linewidth]{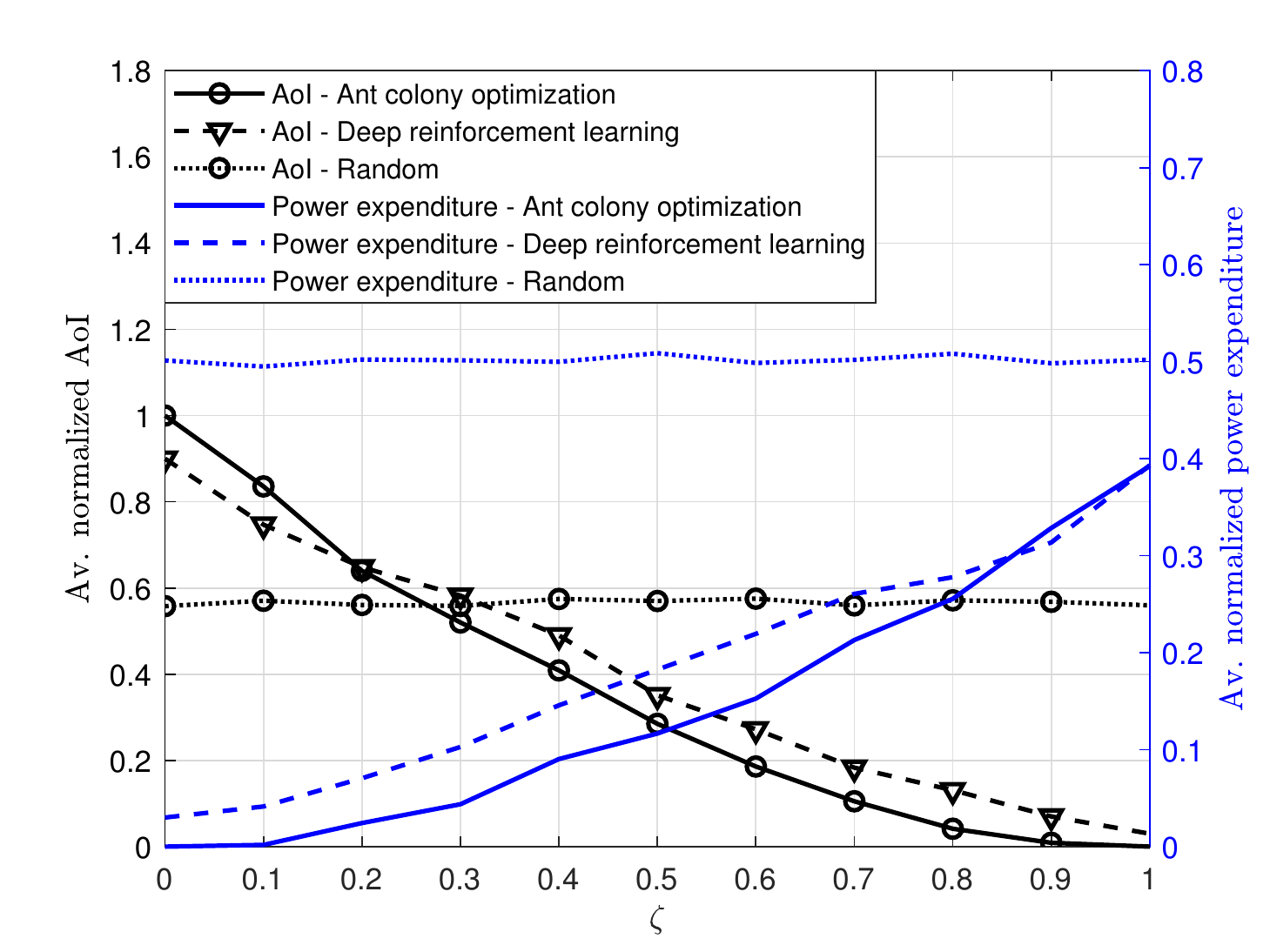}
		\caption{Average normalized  AoI and power expenditure    versus  the relative weight $\zeta$ with total number of processes $F=4$   process and $\abs{\mathcal{R}_i}=2$.}\label{figRes2}
\end{minipage}\hfill
\begin{minipage}{0.301\linewidth}
	\includegraphics[width=1.\linewidth]{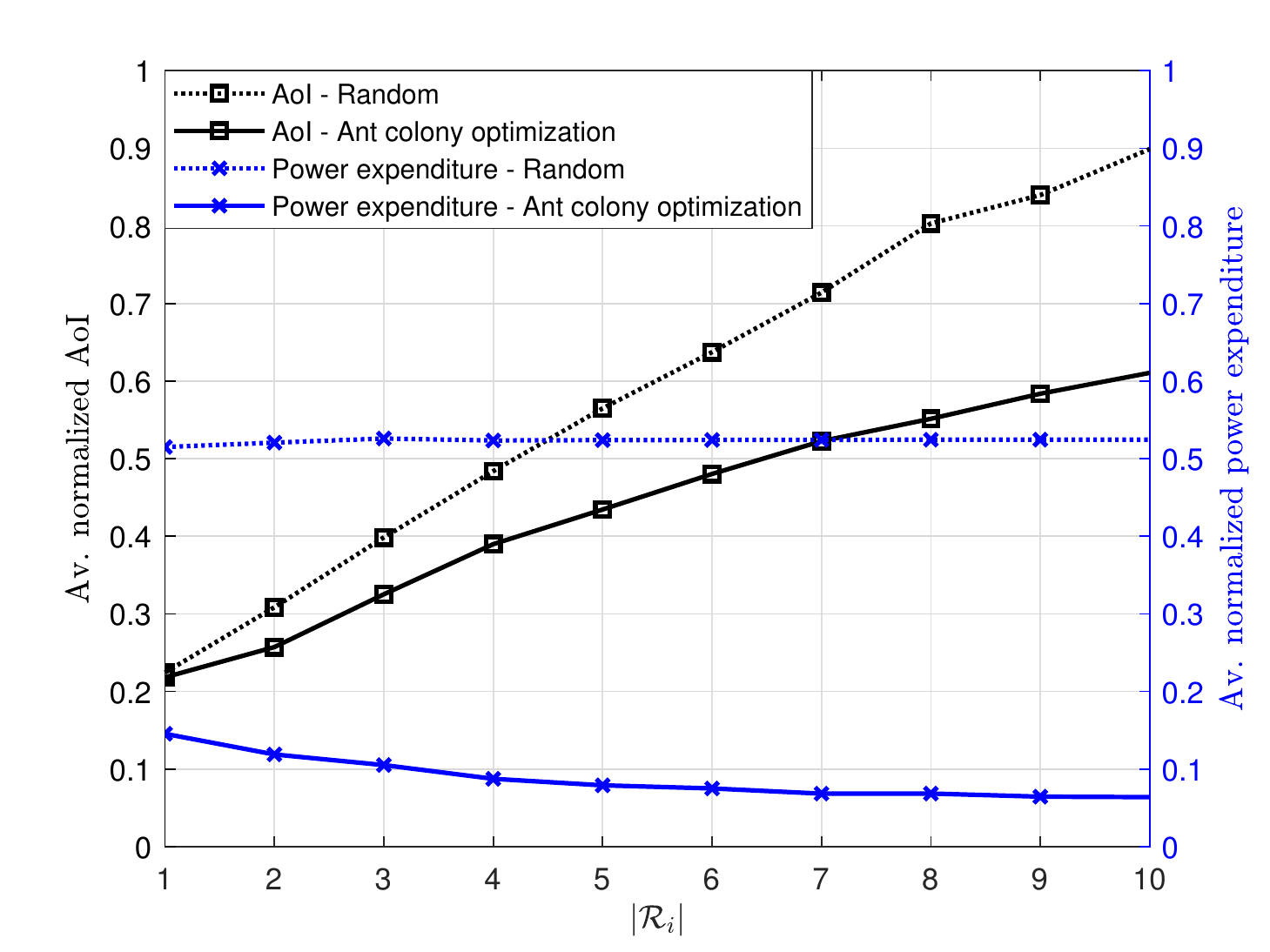}
		\caption{Average normalized  AoI and power expenditure    versus number of process of interest per vehicle $\abs{\mathcal{R}_i}$ with total number of processes $F=10$   process  and $\zeta=0.5$.}\label{figRes3}
\end{minipage}
\label{fig2}
\vspace{-5mm}
\end{figure*}


Figure \ref{figRes1} illustrates the objective function versus the relative weight $ \zeta $ for the proposed framework obtained using the exhaustive search, ACO, and DRL  solution approaches as well as the random approach.  It is seen that the      ACO and DRL approaches achieve performance close to that of the exhaustive search approach and the random approach provides the worst performance in comparison with the proposed framework with the three solution approaches.

To get more insight into this result,  Fig. \ref{figRes2} shows the average normalized AoI and power consumption versus the relative weight $ \zeta $ for the random solution and the   proposed framework using the ACO and DRL solutions. The curves of the exhaustive search   follow a similar trend to that of the ACO and are omitted to make Fig. \ref{figRes2} less crowded.   
It is   noticed that the proposed framework  provides a good trade-off between  AoI and power expenditure as for low values of $ \zeta $ it minimizes the power expenditure and as
$ \zeta $ increases  it minimizes the AoI. That is not the case for the random solution, in which both the AoI and power expenditure are not function of the relative weight and the power consumption is higher than that of the proposed framework.


Figure \ref{figRes3} depicts the average normalized  AoI and power expenditure  of the  random solution  and the proposed framework using the ACO solution   versus the number of process of interest per vehicle $\abs{\mathcal{R}_i}$. It is clear that the AoI increases as the vehicles' demand increases in both random solution and proposed framework, with less AoI in the proposed framework. On the other hand, the  power expenditure in the proposed framework is decreased as  the vehicles' demand increases. This is attributed to   the fact that for a fixed set of processes, as the  number of process of interest per vehicle increases the demand of the vehicles becomes more similar which enables the proposed framework to transmit the same update to more vehicles, which reduces the interference, and thus reduces the  power expenditure.


\section{Conclusion}\label{Con}
	This paper has proposed a  dynamically unicast, multicast, and broadcast transmission framework to minimize both AoI and power consumption in vehicular networks. To solve the formulated mixed integer optimization problem, two solution approaches have been developed, namely a metaheuristic solution  based on ACO and  less computational complex in real-time evaluation solution based on DRL approach.
 Simulation results have illustrated that the proposed framework minimizes both the AoI and power consumption and provides a good trade-off between them. Results also have showed that ACO and DRL solution approaches provide  close to the optimal	solution, which is obtained through exhaustive search.	

\section*{Acknowledgment}
This research was supported by two Discovery Grants funded by the Natural Sciences and Engineering Research Council of Canada.

\bibliographystyle{IEEEtran}
\bibliography{IEEEabrv,Refrences-library}


\end{document}